\def\draftversion{false}

\RequirePackage{ifthen}
\ifthenelse{\equal{\draftversion}{true}}{
  \documentclass[aps,prb,10pt,galley,amsmath,amssymb,showpacs,letterpaper]{revtex4}
}{
  \documentclass[aps,prb,10pt,twocolumn,amsmath,amssymb,showpacs,letterpaper]{revtex4-1}
}

\usepackage{graphicx}
\usepackage{dcolumn}
\usepackage{amssymb}
\usepackage{bm}
\usepackage{enumitem}

\usepackage{epsfig}
\usepackage{hyperref}

\hypersetup{backref=true, pdfnewwindow=true, colorlinks=true,linkcolor=blue, anchorcolor=blue, citecolor=blue, filecolor=blue, menucolor=blue, urlcolor=blue}

\usepackage[usenames,dvipsnames]{color}

\ifthenelse{\equal{\draftversion}{true}}{
  \marginparwidth 2.7in
  \marginparsep 0.5in
  \newcounter{comm} 
  \def\commnext{\stepcounter{comm}}
  \def\commtext{{\bf\color{blue}[\arabic{comm}]}}
  \def\commmar{{\bf\color{blue}[\arabic{comm}]}}
  \def\dvm#1{\commnext\marginpar{\small DV\commmar: #1}\commtext}
  \def\mtm#1{\commnext\marginpar{\small MT\commmar: #1}\commtext}
  \def\jhm#1{\commnext\marginpar{\small JH\commmar: #1}\commtext}
  \def\vsm#1{\commnext\marginpar{\small VS\commmar: #1}\commtext}
  \def\mlab#1{\marginpar{\small\bf #1}}
  
}{
  \def\dvm#1{}
  \def\mtm#1{}
  \def\jhm#1{}
  \def\vsm#1{}
  \def\mlab#1{}
  
}

\begin{document}

\title{First-principles based Landau-Devonshire potential for BiFeO$_3$}

\author{P. Marton}
\email{marton@fzu.cz}
\author{A. Kl\' i\v c}
\author{M. Pa\'{s}ciak}
\author{J. Hlinka}
\affiliation{Institute of Physics, Academy of Sciences of the Czech Republic\\%
Na Slovance 2, 182 21 Prague 8, Czech Republic}
\affiliation{Institute of Mechatronics and Computer Engineering, Technical University of Liberec\\%
Studentsk\'{a} 2, 461 17 Liberec, Czech Republic}

\begin{abstract}

The work describes a first-principles-based computational strategy  for studying structural phase transitions, and in particular, for determination of
the so-called Landau-Devonshire potential  -- the classical zero-temperature limit of the Gibbs energy, expanded in terms of order parameters.
It exploits the  configuration space attached to the eigenvectors of the modes frozen in the ground state, rather than   the space spanned by the unstable modes of the high-symmetry phase, as done usually.
This allows us to carefully probe the part of the energy surface in the vicinity of the ground state, which is most relevant for the properties of the ordered phase.
We apply this procedure to BiFeO$_3$ and perform ab-initio calculations in order to determine
potential energy contributions associated with strain, polarization and oxygen octahedra tilt degrees of freedom, compatible with its two-formula unit cell periodic boundary conditions.

\end{abstract}



\maketitle

Phenomenological models, taking into account important structural order parameters and coupling between them, can greatly help to grasp the physical mechanisms involved in various crystal structure based phenomena, such as piezoelectricity, ferroelectricity, electrostriction etc.
Typically, when simple models for structural phase transitions  are derived from first-principles calculations, the microscopic order parameters relevant for structural phase transition are selected from unstable phonon modes of the high-symmetry reference phase. Then, the configurational space attached to the  Landau-Devonshire (LD) potential (zero-temperature limit of the Gibbs free energy functional) is defined by perturbation of the high-symmetry state along the coordinates associated with these unstable modes.\cite{art_zhong_1994, boo_rabe_2007, art_paul_2017, art_nakhmanson_2010, art_olsen_2016}

Here, we present an alternative approach consisting in the adjustment of the LD potential landscape  in the vicinity of the ground state configuration, which is the most relevant region of the order parameter space when   the ordered phase itself is in the focus of interest. This is particularly important when the disordered phase and ordered ground state have rather distinct atomic and electronic structures. It actually  happens for many materials showing phase transitions at high temperatures.

An interesting material, where such an approach is desirable, is the ferroelectric BiFeO$_3$ -- an insulating  material with a metallic paraelectric phase, known by its large spontaneous polarization and a large $T_{\rm C}$.\cite{Catalan07,Sando14,art_neaton_2005}
The symmetry-breaking order parameters of its ferroelectric $Pm\bar{3}m > R3c$ phase transition, namely
the ferroelectric polarization vector $\bf P$ and the oxygen-octahedron tilt vector $\bf A$, are in our approach associated with the atomic displacement patterns frozen in the fully relaxed ground state. Therefore, the eigenvectors of the high-symmetry phase dynamical matrix are not used here. Actually, in this particular case, the displacement patterns associated with the $\bf P$ and $\bf A$ vectors together with the spontaneous deformation tensor $\bf e$ define the difference between the disordered reference and the ordered ground state completely.

\begin{figure}
\includegraphics[width=8.5cm,clip=true,angle=0]{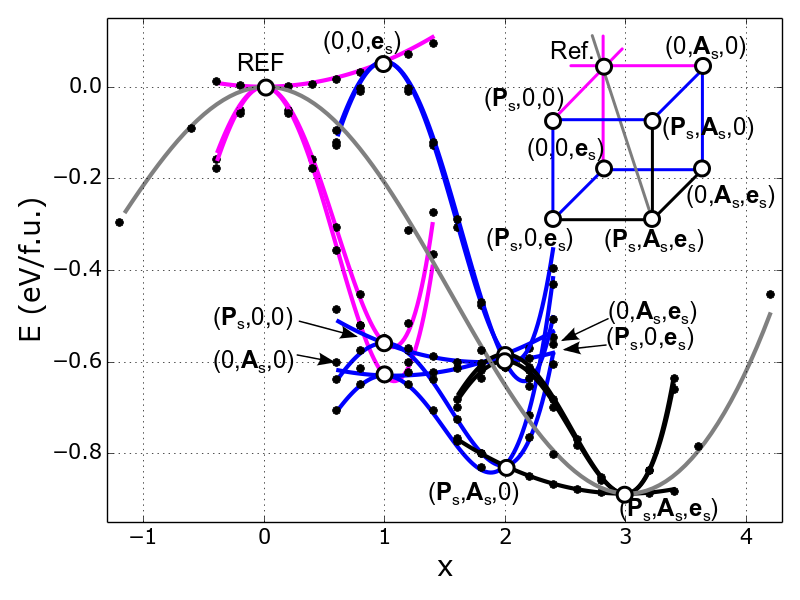}
\caption{
Energy profiles along selected paths in order-parameter space
of BiFeO$_3$, connecting the cubic reference state (REF) and the ground state (${\bf P}_{\rm s},{\bf A}_{\rm s},{\bf e}_{\rm s}$). Point symbols are direct first-principles calculations, lines are evaluated from the present LD potential. Individual path segments are depicted in the inset and described in detail in Table\,I.
}
\label{fig_energy_rhombohedral_distortions}
\end{figure}

The energy profiles along selected paths connecting the reference paraelectric state with  (${\bf P}=0$, ${\bf A}=0$ and ${\bf e}=0$) to the ab-initio ground state (${\bf P}={\bf P}_{\rm s}$, ${\bf A}={\bf A}_{\rm s}$ and ${\bf e}={\bf e}_{\rm s}$) are shown in Fig.\,1. We believe that the numerical values of the LD potential parameters given in this work can be readily used for a variety of purposes such as estimation of nonlinear  electromechanical properties of BiFeO$_3$, influence of the epitaxial strain or in a range of phase-field modeling tasks.

{\em Analytic form of the potential.}
The LD potential (zero-temperature potential density) $f$ considered here
is expanded  around the reference paraelectric state around in the usual form of the symmetry-restricted Taylor expansion in terms of 12 independent variables, covering the selected set of order parameters ${\bf P}=(P_{\rm x},P_{\rm y},P_{\rm z})$, ${\bf A}=(A_{\rm x},A_{\rm y},A_{\rm z})$ and ${\bf e}=(e_{\rm xx},e_{\rm yy},e_{\rm zz},e_{\rm yz},e_{\rm xz},e_{\rm xy})$. Resulting expression can be expressed as a sum
\begin{widetext}
\begin{eqnarray}\label{eqn_total_potential}
f=&& f_{\rm a}^{({\rm e})}[\{P_i\}]+ f_{\rm b}^{({\rm e})}[\{A_i\}]+ f_{\rm t}^{({\rm e})}[\{P_i,A_i\}]+f_{\rm est}[\{P_i,e_{ij}\}]+f_{\rm rst}[\{A_i,e_{ij}\}]+f_{\rm ela}[\{e_{ij}\}] ~.
\end{eqnarray}
The first three terms describe the energy expansion in term of the ferroelectric polarization and oxygen octahedron tilt only. They have been expanded till 8-th order because stopping at the 6-th order did not reproduced the ab-initio calculated potential landscape satisfactorily.
The ferroelectric part of the Landau energy, $f_{\rm a}^{\rm ({\rm e})}$,
contains all symmetry allowed terms
\begin{eqnarray}
\label{eqn_form_P}
f_{\rm a}^{\rm ({\rm e})}=
  a_{i} P_{i}^2
+ a_{ij}^{\rm ({\rm e})}P_{i}^2P_{j}^2
+ a_{ijk}P_{i}^2P_{j}^2P_{k}^2
+ a_{ijkl}P_{i}^2P_{j}^2P_{k}^2P_{l}^2~,
\end{eqnarray}
and the same holds for the $f_{\rm b}^{\rm ({\rm e})}$, which has the same form but in terms of angles $\{A_i\}$ and coefficients $b$. In case of the coupling between polarization and oxygen octahedron tilt coupling terms
\begin{eqnarray}
\label{eqn_form_PA}
f_{\rm t}^{({\rm e})}=
  t_{ijkl}^{\rm ({\rm e})}P_{i}P_{j} A_{k}A_{l}
+ t_{ijklmn}^{42} P_{i}P_{j}P_{k}P_{l} A_{m}A_{n}
+ t_{ijklmn}^{24} P_{i}P_{j}A_{k}A_{l}A_{m} A_{n}
+ t_{ijklmnpq}^{44}P_{i}P_{j}P_{k}P_{l} A_{m} A_{n}A_{p} A_{q}\\\nonumber
+ t_{ijklmnpq}^{62}P_{i}P_{j}P_{k}P_{l} P_{m} P_{n}A_{p} A_{q}
+ t_{ijklmnpq}^{26}P_{i}P_{j}A_{k}A_{l} A_{m} A_{n}A_{p} A_{q}~,
\end{eqnarray}
 we have included all symmetry allowed terms up to the 6-th order, while the 8-th order terms were limited to the pair interaction between  single $P_i$ and single $A_i$ components only. On the other hand, since the strain contribution to the ground state LD energy are rather small, only the lowest order coupling and self-energy terms in strain were introduced. Their role is to describe electrostriction, rotostriction (introduced in an obvious analogy to electrostriction), and elastic energy contributions, respectively:
\begin{eqnarray}
\label{eqn_form_est_rot_ela}
f_{\rm est}=-q_{ijkl}e_{ij} P_{k}P_{l},~~~~~~~~f_{\rm rot}=-r_{ijkl}e_{ij} A_{k}A_{l},~~~~~~~~f_{\rm ela}=\frac{1}{2}C_{ijkl}e_{ij}e_{kl}~.
\end{eqnarray}
\end{widetext}
Let us note that the term including all three order parameters is not included 
and that the adopted potential form
allows a straightforward analytical elimination of the strain degree of freedom using linear equations of mechanical equilibrium,\cite{art_nambu_sagala_1994,art_hlinka_marton_prb_2006} which facilitates considerations
about a mechanically free crystal.

The superscript (e) marks the terms and parameters which are renormalized upon such strain elimination, and it emphasizes that they are related to material clamped to the reference cubic shape.

{\em First-principles-calculations details.}
The total-energy calculations are based on density functional theory (DFT) within the local spin-density approximation (LSDA) method as implemented in the Vienna {\em ab-initio} Simulation Package (VASP).\cite{art_kresse_hafner_1993, art_kresse_furthmuller_1996}
The projector-augmented plane-wave method was used.\cite{art_blochl_1994}
There were 15 explicitly treated electrons for Bi (5d$^{10}$\,6s$^2$\,6p$^3$), 14 for Fe (3p$^6$\,4s$^1$\,3d$^7$) and six for oxygen (2s$^2$\,2p$^4$).
Energy cutoff for plane-waves was set to 500\,eV.
The Brillouin-zone integrations were carried out using $3\,\times\,3\,\times\,3$ Monkhorst-Pack $k$-point mesh.\cite{art_monkhorst_pack_1976}
Gaussian broadening of 0.1\,eV was applied.\cite{art_fu_1983}
The antiferromagnetic (AFM) G-type order on Fe atoms has been taken into account.
%
%
The LSDA+U method is applied with the Hubbard term\cite{art_anisimov_1997} added to the iron $d$-orbitals.
In the used Dudarev\cite{art_dudarev_1998} approach the difference $U-J$ was set to 3.0\,eV.

All calculations presented here were performed using a 10-atom supercell of BiFeO$_3$
compatible with the G-type AFM order.
The two perovskite cells in the supercell host two oppositely rotated oxygen octahedra.
The tilt vector ${\bf A}$ discussed in the text always refers to the axial rotation vector of the oxygen octahedron in the first cell, while the rotation in the other cell is automatically opposite (-${\bf A}$).
Polarization vector ${\bf P}$ is the same in both perovskite cells.

{\em Microscopic content of the selected order parameters.}
As a starting point in the presented procedure the atomic structure of the reference cubic and the rhombohedral ground states of the BiFeO$_3$ were determined.
They are fully optimized using first-principles calculations within cubic and $R3c$ symmetries, respectively, leading to configurations in good agreement with the available theoretical\cite{art_graf_2014,art_neaton_2005} and experimental\cite{art_arnold_2009} data.

The difference between the ground-state and the reference configurations can be specified by differences in fractional atomic coordinates  $\Delta{\bf u}_{\rm s}$  and by the change of the supercell lattice vectors.
Evaluation of the strain tensor is a straightforward procedure, because the lattice vectors of any slightly distorted unit cell described by the tensor of deformation ${\bf e}$ can be obtained by applying the ${\bf 1}+{\bf e}$ matrix multiplication operation to the three reference-state lattice vectors.
However, the $\Delta{\bf u}_{\rm s}$ (which is a 30-component vector for the ten-atom BiFeO$_3$ supercell) requires more attention. It can be further decomposed into two parts, $\Delta{\bf u}_{\rm s}^{(a)} + \Delta{\bf u}_{\rm s}^{(b)}$ .
The first part, which transforms as the (111) component of the $F_{1u}$ Brillouin zone center polar mode of the parent cubic phase, defines displacements $\Delta{\bf u}_{\rm s}^{(a)}$ related to the spontaneous ferroelectric polarization ${\bf P}_{\rm s} \parallel (111)$, while
the remaining part, $\Delta{\bf u}_{\rm s}^{(b)}$, transforming as the Brillouin zone corner irreducible representation, defines the spontaneous oxygen octahedron tilt ${\bf A}_{\rm s} \parallel (111)$.
This decomposition is thus unique.

Furthermore, we have assumed that the space of vectors ${\bf P}$ are attached to a  single zone-center mode, {\it i.e.} to a mode with a fixed atomic pattern. In other words, the atomic displacements corresponding to the polarization ${\bf P}=\xi{\bf P}_{\rm s}$ are given by $\Delta{\bf u}^{(a)}=\xi \Delta{\bf u}_{\rm s}^{(a)}$ and the atomic displacements corresponding to the polarization ${\bf P}$  equal to the spontaneous one but directed along, say (100) Cartesian direction, are obtained by rotation of each individual atomic displacement vector comprised in $\Delta{\bf u}_{\rm s}^{(a)}$ by the same proper rotation that turns (111) into  (100) direction in the space of the attached polarization vectors ${\bf P}$.
Similar procedure is applied to relate $\Delta{\bf u}_{\rm s}^{(b)}$
with ${\bf A}$. This construction defines a consistent linear subspace of the atomic coordinates compatible with the $Z=2$ supercell boundary conditions.

\begin{table}
\begin{tabular}{lllllr}
\hline \hline
~~~~&\multicolumn{1}{c}{${ P_i}/(P_{\rm s})_i$}&\multicolumn{1}{c}{${ A_i}/(A_{\rm s})_i$}&\multicolumn{1}{c}{${  e}_{ij}/({  e}_{\rm s})_{ij}$}&x\\
\hline
 1~~~~~~~~~~~~&($\xi$,$\xi$,$\xi$,~~~~~~~~&     	0,0,0,~~~~~~~~&     		0,0,0,0,0,0)~~~~~~~~&			$\xi$\\
 2&(0,0,0,&     	$\xi$,$\xi$,$\xi$,&     	0,0,0,0,0,0)&			$\xi$\\
 3&(0,0,0,&     	0,0,0,&    		$\xi$,$\xi$,$\xi$,$\xi$,$\xi$,$\xi$)&	$\xi$\\
 4&(1,1,1,&     	$\xi$,$\xi$,$\xi$,&     	0,0,0,0,0,0)&			1+$\xi$\\
 5&(1,1,1,&     	0,0,0,&    		 $\xi$,$\xi$,$\xi$,$\xi$,$\xi$,$\xi$)&	1+$\xi$\\
 6&($\xi$,$\xi$,$\xi$,&    	1,1,1,&     		0,0,0,0,0,0)&			1+$\xi$\\
 7&(0,0,0,&     	1,1,1,&     		$\xi$,$\xi$,$\xi$,$\xi$,$\xi$,$\xi$)&	1+$\xi$\\
 8&($\xi$,$\xi$,$\xi$,&     	0,0,0,&     		1,1,1,1,1,1)&			1+$\xi$\\
 9&(0,0,0,&     	$\xi$,$\xi$,$\xi$,&     	1,1,1,1,1,1)&			1+$\xi$\\
10&($\xi$,$\xi$,$\xi$,&     	1,1,1,&     		1,1,1,1,1,1)&			2+$\xi$\\
11&(1,1,1,&     	$\xi$,$\xi$,$\xi$,&     	1,1,1,1,1,1)&			2+$\xi$\\
12&(1,1,1,&     	1,1,1,&     		$\xi$,$\xi$,$\xi$,$\xi$,$\xi$,$\xi$)&	2+$\xi$\\
13&($\xi$,$\xi$,$\xi$,&     	$\xi$,$\xi$,$\xi$,&     	$\xi$,$\xi$,$\xi$,$\xi$,$\xi$,$\xi$)&	3$\xi$\\
\hline \hline
\end{tabular}
\caption{
Paths in the order-parameter space, sampled by a
path parameter $\xi$, for which the first-principles energies were evaluated and plotted in Fig.\,\ref{fig_energy_rhombohedral_distortions}.
The last column corresponds to the position of the calculated point on the horizontal axis in the Fig\,\ref{fig_energy_rhombohedral_distortions}.
}
\label{tab_paths_rhombohedral_distortions}
\end{table}

\begin{table}
\begin{tabular}{llrrll}
\hline \hline
~&Type&~~&&SI&~~Unit\\
\hline
$a_{1}$		&$P_{\rm x}^2$						&$-3.362$&$\times$&$ 10^{9}$		&J\,m\,C$^{-2}$\\
$a_{11}^{(\rm e)}$	&$P_{\rm x}^4$						&$2.646$&$\times$&$ 10^{9}$		&J\,m\,C$^{-2}$\\
$2a_{12}^{(\rm e)}$	&$P_{\rm x}^2P_{\rm y}^2$				&$3.274$&$\times$&$ 10^{9}$		&J\,m\,C$^{-2}$\\
$a_{111}$		&$P_{\rm x}^6$						&$-5.960$&$\times$&$ 10^{8}$		&J\,m\,C$^{-2}$\\
$3a_{112}$		&$P_{\rm x}^4P_{\rm y}^2$				&$2.634$&$\times$&$ 10^{8}$		&J\,m\,C$^{-2}$\\
$6a_{123}$		&$P_{\rm x}^2P_{\rm y}^2P_{\rm z}^2$			&$-7.132$&$\times$&$ 10^{9}$		&J\,m\,C$^{-2}$\\
$a_{1111}$		&$P_{\rm x}^8$						&$9.043$&$\times$&$ 10^{7}$		&J\,m\,C$^{-2}$\\
$4a_{1112}$		&$P_{\rm x}^6P_{\rm y}^2$				&$-2.284$&$\times$&$ 10^{8}$		&J\,m\,C$^{-2}$\\
$6a_{1122}$		&$P_{\rm x}^4P_{\rm y}^4$				&$4.636$&$\times$&$ 10^{8}$		&J\,m\,C$^{-2}$\\
$12a_{1123}$		&$P_{\rm x}^4P_{\rm y}^2P_{\rm z}^2$			&$1.493$&$\times$&$ 10^{9}$		&J\,m\,C$^{-2}$\\
$b_{1}$		&$A_{\rm x}^2$						&$-1.585$&$\times$&$ 10^{7}$		&J\,m$^{-3}$\,deg.$^{-2}$\\
$b_{11}^{(\rm e)}$	&$A_{\rm x}^4$						&$5.396$&$\times$&$ 10^{4}$		&J\,m$^{-3}$\,deg.$^{-2}$\\
$2b_{12}^{(\rm e)}$	&$A_{\rm x}^2A_{\rm y}^2$				&$6.314$&$\times$&$ 10^{4}$		&J\,m$^{-3}$\,deg.$^{-2}$\\
$b_{111}$		&$A_{\rm x}^6$						&$-6.598$&$\times$&$ 10^{1}$		&J\,m$^{-3}$\,deg.$^{-2}$\\
$3b_{112}$		&$A_{\rm x}^4A_{\rm y}^2$				&$-5.203$&$\times$&$ 10^{1}$		&J\,m$^{-3}$\,deg.$^{-2}$\\
$6b_{123}$		&$A_{\rm x}^2A_{\rm y}^2A_{\rm z}^2$			&$-5.910$&$\times$&$ 10^{1}$		&J\,m$^{-3}$\,deg.$^{-2}$\\
$b_{1111}$		&$A_{\rm x}^8$						&$4.890$&$\times$&$ 10^{-2}$		&J\,m$^{-3}$\,deg.$^{-2}$\\
$4b_{1112}$		&$A_{\rm x}^6A_{\rm y}^2$				&$1.598$&$\times$&$ 10^{-2}$		&J\,m$^{-3}$\,deg.$^{-2}$\\
$6b_{1122}$		&$A_{\rm x}^4A_{\rm y}^4$				&$1.194$&$\times$&$ 10^{-1}$		&J\,m$^{-3}$\,deg.$^{-2}$\\
$12b_{1123}$		&$A_{\rm x}^4A_{\rm y}^2A_{\rm z}^2$			&$1.002$&$\times$&$ 10^{-1}$		&J\,m$^{-3}$\,deg.$^{-2}$\\
$t^{(\rm e)}_{1111}$	&$P_{\rm x}^2A_{\rm x}^2$				&$1.720$&$\times$&$ 10^{7}$		&J\,m\,C$^{-2}$\,deg.$^{-2}$\\
$t^{(\rm e)}_{1122}$	&$P_{\rm x}^2A_{\rm y}^2$				&$2.273$&$\times$&$ 10^{7}$		&J\,m\,C$^{-2}$\,deg.$^{-2}$\\
$4t^{(\rm e)}_{1212}$	&$P_{\rm x}P_{\rm y}A_{\rm x}A_{\rm y}$			&$-2.844$&$\times$&$ 10^{7}$		&J\,m\,C$^{-2}$\,deg.$^{-2}$\\
$t^{42}_{111111}$	&$P_{\rm x}^4A_{\rm x}^2$				&$2.371$&$\times$&$ 10^{6}$		&J\,m$^5$\,C$^{-4}$\,deg.$^{-2}$\\
$t^{24}_{111111}$	&$P_{\rm x}^2A_{\rm x}^4$				&$-5.689$&$\times$&$ 10^{4}$		&J\,m\,C$^{-2}$\,deg.$^{-4}$\\
$t^{42}_{111122}$	&$P_{\rm x}^4A_{\rm y}^2$				&$-9.069$&$\times$&$ 10^{6}$		&J\,m$^5$\,C$^{-4}$\,deg.$^{-2}$\\
$t^{24}_{112222}$	&$P_{\rm x}^2A_{\rm y}^4$				&$-4.608$&$\times$&$ 10^{4}$		&J\,m\,C$^{-2}$\,deg.$^{-4}$\\
$6t^{42}_{112233}$	&$P_{\rm x}^2P_{\rm y}^2A_{\rm z}^2$			&$-8.438$&$\times$&$ 10^{6}$		&J\,m$^5$\,C$^{-4}$\,deg.$^{-2}$\\
$6t^{24}_{112233}$	&$P_{\rm x}^2A_{\rm y}^2A_{\rm z}^2$			&$-2.421$&$\times$&$ 10^{4}$		&J\,m\,C$^{-2}$\,deg.$^{-4}$\\
$6t^{42}_{112211}$	&$P_{\rm x}^2P_{\rm y}^2A_{\rm x}^2$			&$6.805$&$\times$&$ 10^{6}$		&J\,m$^5$\,C$^{-4}$\,deg.$^{-2}$\\
$6t^{24}_{111122}$	&$P_{\rm x}^2A_{\rm x}^2A_{\rm y}^2$			&$2.594$&$\times$&$ 10^{4}$		&J\,m\,C$^{-2}$\,deg.$^{-4}$\\
$8t^{42}_{111212}$	&$P_{\rm x}^3P_{\rm y}A_{\rm x}A_{\rm y}^2$		&$-1.954$&$\times$&$ 10^{7}$		&J\,m$^5$\,C$^{-4}$\,deg.$^{-2}$\\
$24t^{42}_{123312}$	&$P_{\rm x}P_{\rm y}P_{\rm z}^2A_{\rm x}A_{\rm y}$	&$2.660$&$\times$&$ 10^{7}$		&J\,m$^5$\,C$^{-4}$\,deg.$^{-2}$\\
$8t^{24}_{121112}$	&$P_{\rm x}P_{\rm y}A_{\rm x}^3A_{\rm y}$		&$-8.314$&$\times$&$ 10^{3}$		&J\,m\,C$^{-2}$\,deg.$^{-4}$\\
$24t^{24}_{121233}$	&$P_{\rm x}P_{\rm y}A_{\rm x}A_{\rm y}A_{\rm z}^2$	&$-5.457$&$\times$&$ 10^{4}$		&J\,m\,C$^{-2}$\,deg.$^{-4}$\\
$t^{62}_{11111111}$	&$P_{\rm x}^6A_{\rm x}^2$				&$-2.573$&$\times$&$ 10^{6}$		&J\,m$^9$\,C$^{-6}$\,deg.$^{-2}$\\
$t^{26}_{11111111}$	&$P_{\rm x}^2A_{\rm x}^6$				&$1.132$&$\times$&$ 10^{2}$		&J\,m\,C$^{-2}$\,deg.$^{-6}$\\
$t^{62}_{11111122}$	&$P_{\rm x}^6A_{\rm y}^2$				&$1.370$&$\times$&$ 10^{6}$		&J\,m$^9$\,C$^{-6}$\,deg.$^{-2}$\\
$t^{26}_{11222222}$	&$P_{\rm x}^2A_{\rm y}^6$				&$5.255$&$\times$&$ 10^{1}$		&J\,m\,C$^{-2}$\,deg.$^{-6}$\\
$t^{44}_{11111111}$	&$P_{\rm x}^4A_{\rm x}^4$				&$4.747$&$\times$&$ 10^{4}$		&J\,m$^5$\,C$^{-4}$\,deg.$^{-4}$\\
$t^{44}_{11112222}$	&$P_{\rm x}^4A_{\rm y}^4$				&$1.888$&$\times$&$ 10^{4}$		&J\,m$^5$\,C$^{-4}$\,deg.$^{-4}$\\
$q_{1111}$		&$e_{\rm xx}P_{\rm x}^2$				&$1.447$&$\times$&$ 10^{10}$		&J\,m\,C$^{-2}$\\
$q_{1122}$		&$e_{\rm xx}P_{\rm y}^2$				&$4.776$&$\times$&$ 10^{9}$		&J\,m\,C$^{-2}$\\
$2q_{1212}$		&$e_{\rm xy}P_{\rm x}P_{\rm y}$				&$7.186$&$\times$&$ 10^{9}$		&J\,m\,C$^{-2}$\\
$r_{1111}$		&$e_{\rm xx}A_{\rm x}^2$				&$2.319$&$\times$&$ 10^{7}$		&J\,m$^{-3}$\,deg.$^{-2}$\\
$r_{1122}$		&$e_{\rm xx}A_{\rm y}^2$				&$-4.886$&$\times$&$ 10^{6}$		&J\,m$^{-3}$\,deg.$^{-2}$\\
$2r_{1212}$		&$e_{\rm xy}A_{\rm x}A_{\rm y}$				&$-2.526$&$\times$&$ 10^{7}$		&J\,m$^{-3}$\,deg.$^{-2}$\\
$C_{1111}$		&$e_{\rm xx}^2$						&$2.666$&$\times$&$ 10^{11}$		&Pa\\
$C_{1122}$		&$e_{\rm xx}e_{\rm yy}$					&$1.435$&$\times$&$ 10^{11}$		&Pa\\
$C_{1212}$		&$e_{\rm xy}^2$						&$9.548$&$\times$&$ 10^{10}$		&Pa\\
\hline \hline
\end{tabular}
\caption{Coefficients of the LD potential for BiFeO$_3$. Second column provide one representant of symmetry-equivalent terms associated with the respective coefficient, third column gives the numerical value of the coefficient or its simple multiple, indicated in the first column.
}
\label{tab_summary_of_parameters}
\end{table}

{\em Fitting parameters of the Landau-Devonshire potential.}
The described procedure establishes a one-to-one correspondence between point in the order parameters space and atomic structure.
Using this link, it is possible to sample the parameter space, use first-principles calculations to determine energies, and fit the energy surface in order to obtain parameters of the potential.

The individual terms in (\ref{eqn_total_potential}) are fitted separately.
For each term, a suitable set of paths was chosen.
For example, to fit the parameters $\alpha$'s in the $f_{\rm a}^{\rm ({\rm e})}(\{P_i\})$ several paths along the high symmetry directions in the polarization space were chosen, while keeping the oxygen octahedra tilt and strain zero.
Angular dependence of energy on polarization vector was probed employing circular paths around the paraelectric reference state with several different diameters and orientations of the circles.
Similar procedure was then adopted for  $f_{\rm b}^{\rm ({\rm e})}(\{A_i\})$.
In evaluation of the coupling terms, e.g. between polarization and tilt, we subtract the already determined contribution of polarization and tilt and fit only the energy difference corresponding to the coupling energy.
The so obtained set of parameters is consequently utilized as an initial condition for fitting of all parameters together under an additional  constraint ensuring that the position of the global minimum of the potential  corresponds exactly to the ground state obtained from first-principles. This fitting procedure has been conveniently accomplished in dimensionless variables normalized to the ground state values of $\Delta{\bf u}_{\rm s}^{(a)}$, $\Delta{\bf u}_{\rm s}^{(b)}$ and ${\bf e}_{\rm s}$.
The final set of the numerical values of defining the LD potential parameters of BiFeO$_3$ were rescaled with spontaneous values $({\bf e}_{\rm s})_{\rm xx}=0.136$, $({\bf e}_{\rm s})_{\rm xy}=0.0012$, ${ A}_{\rm s} =|{\bf A}_{\rm s} |= 14.355$\, degree (our ab-initio data) and ${ P}_{\rm s} =|{\bf P}_{\rm s} |= 0.91$\,C/m$^2$ (adopted from Ref.\,\onlinecite{art_dieguez_2011}). Resulting Landau-Devonshire model for BiFeO$_3$ is given in Tab.\,\ref{tab_summary_of_parameters}.

\begin{figure}
\includegraphics[width=9.5cm,clip=true,angle=0]{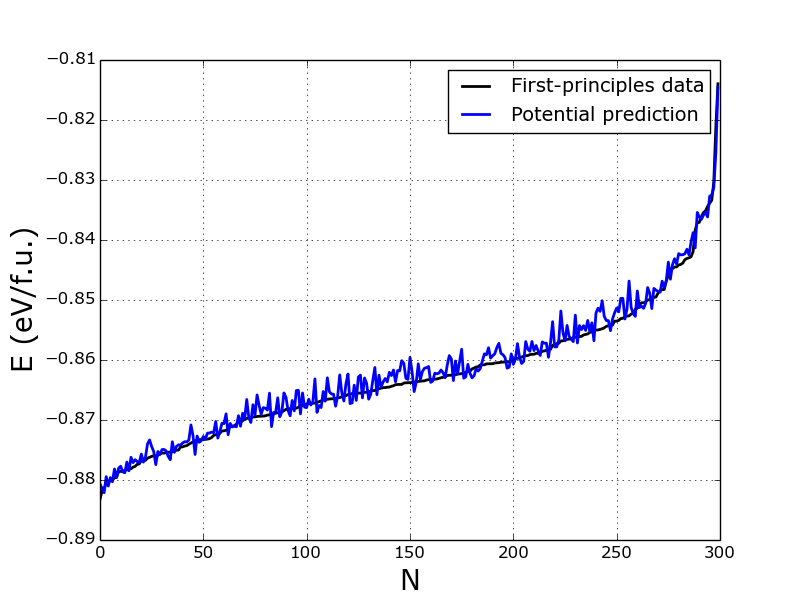}
\caption{
Comparison of first-principles and LD-predicted energy levels for 300 randomly selected points in the order-parameters space in the vicinity of the rhombohedral ground-state (coordinates falling within 20 percent around the ground state values). Monotoneously increasing function stands for the direct first-principles data, ordered by the increasing total energy, while the fluctuating curve connects the corresponding energies, calculated from the LD potential.
}
\label{fig_random_v2}
\end{figure}

The energy profiles of the LD potential along several important directions in the order-parameter space (path listed in Tab.\,\ref{tab_paths_rhombohedral_distortions}) are displayed in Fig.\,\ref{fig_energy_rhombohedral_distortions}).
The direct ab-initio calculated energies, presented as dots in the figure, exhibit visually perfect agreement with the predictions of the fitted LD potential.
Therefore, it can be expected that the present potential also describes well
the various linear response properties, in particular within its ground state.
For example, the elastic tensor of BiFeO$_3$ predicted by second order strain derivatives of the present LD potential in its rhombohedral groundstate compares fairly well with other literature data (see Tab.\,\ref{tab_elastic_coefficients}).
As an even more representative performance test, we have generated 300 random configurations with the 12 order parameter components falling within $\pm 20$ percent around their spontanous values and calculated their total energies from LD potential as well as from the first principles. The agreement is also very satisfactory (see Fig.\,2). Moreover, since the present potential goes well behind the so far adopted  quartic anharmonicity in the polarization and tilt degrees of freedom\cite{Xue2014, Kornev2007, Daraktchiev2010, Kulagin2015, Kornev2006}, one can expect that the this LD potential will be more appropriate when dealing with nonlinear responses.

\begin{table}
\begin{tabular}{llllllll}
\hline \hline
					&$C_{11}$	&$C_{12}$	&$C_{13}$	&$C_{14}$	&$C_{33}$	&$C_{44}$	&$C_{66}$	\\
                    \hline
LD potential (this work)				&278		&122		&95		&-22		&228		&57		&78		\\
LDA	calculations (this work)		&264		&147		&63		&-16		&132		&53		&54		\\
Borissenko et {\em al.}\cite{Borissenko2013}, LDA	&249		&151		&75		&9		&160		&44		&49		\\
Shang et {\em al.}\cite{Shang2009}, GGA	&222		&110		&50		&16		&150		&49		&56		\\
\hline \hline
\end{tabular}
\caption{
Elastic stiffness constants of rhombohedral BiFeO$_3$ ground state (sign of $C_{14}$ depends on adopted coordinate system).
}
\label{tab_elastic_coefficients}
\end{table}

To conclude, we present a comprehensive and efficient
procedure for extraction of the Landau-Devonshire-type potential
from quantum-mechanical calculations. We believe that this scheme, based on the microscopic content of order parameters derived from the full amplitude distortions of the low-symmetry phase ground states,
 will enable real methodological progress for systematic development of models for a large family of materials with structural phase transitions.
In the specific case of the prototypical multiferroic material BiFeO$_3$, we present a carefully engineered  Landau-Devonshire potential, which can be readily used for any analytical or computer simulations requiring its realistic phenomenological LD model.
We believe that the potential and its applications will improve understanding of BiFeO$_3$ as well as open an avenue for computationally supported engineering of BiFeO$_3$-based functional structures.

This work was supported by the Czech Science Foundation (project No.\,15-04121S). Access to computing and storage facilities owned by parties and projects contributing to the National Grid Infrastructure MetaCentrum, provided under the program "Projects of Large Research, Development, and Innovations Infrastructures" (CESNET LM2015042), is greatly appreciated.

\bibliographystyle{unsrtnat}            	

\end{document}